\documentclass{aastex}
\usepackage{emulateapj5}
\usepackage{apjfonts}

\usepackage{natbib}
\usepackage{epsf}
\usepackage{psfig}

\shorttitle{Internal kinematics of distant cluster spirals}
\shortauthors{Ziegler et al.}

\newcommand{\clvier}{Cl\,0413--65}
\newcommand{\mseins}{MS\,1008--12}
\newcommand{\cldrei}{Cl\,0303$+$17}
\newcommand{\datapaper}{Paper\,II}

\received{2003 July 11}
\begin{document}

\bibliographystyle{abbrvnat}

\title{Internal kinematics of spiral galaxies in distant clusters Part\,I
\altaffilmark{1}}
\altaffiltext{1}{Based on observations collected at the European
Southern Observatory, Cerro Paranal, Chile (ESO Nos. 64.O--0158, 64.O--0152 
\& 66.A--0547)}

\author{
B. L. Ziegler, A. B\"ohm, K. J\"ager}
\affil{Universit\"ats--Sternwarte G\"ottingen, Geismarlandstr. 11, 37083
G\"ottingen, Germany}
\email{bziegler@uni-sw.gwdg.de, jaeger@uni-sw.gwdg.de, boehm@uni-sw.gwdg.de}
\and
\author{J. Heidt, C.~M\"ollenhoff}
\affil{Landessternwarte Heidelberg, K\"onigstuhl, D--69117 Heidelberg, Germany}
\email{J.Heidt@lsw.uni-heidelberg.de, C.Moellenhoff@lsw.uni-heidelberg.de}


\begin{abstract}
We introduce our project on galaxy evolution in the environment of rich clusters 
aiming at disentangling the importance of specific interaction and galaxy 
transformation processes from the hierarchical evolution of galaxies in the field.
Emphasis is laid on the examination of the internal kinematics of disk galaxies 
through spatially resolved MOS spectroscopy with FORS at the VLT. First results 
are presented for the clusters MS\,1008.1--1224 ($z=0.30$), Cl\,0303$+$1706
($z=0.42$), and Cl\,0413--6559 (F1557.19TC) ($z=0.51$). 
Out of 30 cluster members with emission-lines, 13 galaxies exhibit a rotation 
curve of the universal form rising in the inner region and passing over into a 
flat part. The other members have either intrinsically peculiar kinematics (4), or
too strong geometric distortions (9) or too low $S/N$ (4 galaxies) for a reliable
classification of their velocity profiles.
The 13 cluster galaxies for which a maximum rotation velocity could be derived 
are distributed in the Tully--Fisher diagram very similar to field 
galaxies from the FORS Deep Field that have corresponding redshifts and do not
show any significant luminosity evolution with respect to local samples.
The same is true for seven galaxies observed in the cluster fields that turned out 
not to be members.
The mass--to--light ratios of the 13 TF cluster spirals cover the
same range as the distant field population indicating that their stellar 
populations were not dramatically changed by possible clusterspecific interaction
phenomena.
The cluster members with distorted kinematics may be subject to interaction processes
but it is impossible to determine whether these processes also lead to changes in the 
overall luminosity of their stellar populations.
\end{abstract}



\section{Introduction}

Galaxy clusters provide a special environment for their members. In
contrast to the field, the number volume density of galaxies is high and the
relative velocities are large. The gravitational potential of a cluster is
filled by the intracluster medium (ICM), a hot X--ray emitting gas, and the
overall mass--to--light ratio is much larger than for the individual galaxies
indicating the presence of vast amounts of dark matter. This environment exerts
a strong influence on the evolution of the cluster galaxies superposed on the
(field) evolution that arises from the hierarchical growth of objects
and the declining starformation rates over cosmic epochs.
Besides tidal interactions between galaxies including merging as can also be 
observed in the field, cluster members are affected by clusterspecific 
phenomena related to the ICM (like ram-pressure stripping) or the structure of 
the cluster (like harassment). For a recent overview see third volume of the
Carnegie Observatories Astrophysics Series\footnote{
http://www.ociw.edu/ociw/symposia/series/symposium3/proceedings.html}.
Imprints of these
interactions can not only be seen in present-day clusters, but also manifest
themselves in a strong evolution of the population of cluster galaxies.
One example is the photometric Butcher--Oemler effect of an increasing fraction 
of blue galaxies with redshift \citep[e.g.][]{BO78b} implying a rising
percentage of starforming galaxies. Another example is the rapid decline of the 
abundance of lenticular galaxies (S0) from the dominant population in local 
clusters to a few percent at a lookback time of $\sim5$\,Gyrs 
\citep[e.g.][]{DOCSE97}.
These observations have led to the question whether field spirals falling
into a cluster can be subject to such morphological transformations that they
appear as S0 galaxies today. 

Independent from whether this overall scenario is true or not, the observed
tidal interactions (either between galaxies or with the cluster potential) 
may cause substantial distortions both on the structure and the kinematics of
the galaxies involved. Indeed, \citet{RWK99}, for example, found that half of
their sample of 89 disk galaxies in the Virgo cluster exhibit kinematic
disturbances ranging from modest (e.g. asymmetric) to severe (e.g. truncated 
curves) peculiarities. 
On the other hand, many local cluster galaxies, for which only H\,{\sc i} 
velocity widths (in 
contrast to spatially resolved velocity profiles) were measured, follow a 
tight Tully--Fisher relation similar to field spirals \citep[e.g.][]{GHHVC97}.
This TFR connects the luminosity of the stellar population of a galaxy to its
internal kinematics which are dominated by the presence of a dark matter halo
\citep{TF77}.

But it is not yet clear whether the halo of dark matter and, therefore, the
total mass of a galaxy can also be affected by certain interaction phenomena. 
In numerical simulations of the evolution of substructure in clusters by
\citet{SWTK01}, the dark matter halo of a galaxy that falls into the cluster
is truncated via tidal interactions so that the mass-to-light ratio of a galaxy
gets reduced during its passage to the cluster core.
\citet{Gnedi03b} simulates the tidal field along galactic orbits in
hierarchically growing clusters, and finds that about 40\% of the dark halo 
of a
massive galaxy ($V_{\rm max}=250$\,km/s) is lost between $z=5$ and $z=0$. But
the rotational velocity at $\sim5$ disk scale lengths is predicted to hardly 
change (decrease by $\sim2$\%).

\section{Our Project}

Since in models of hierarchically growing structure clusters are still in the 
process of forming at $z\lesssim1$ in the concordance cosmology, a higher 
infall rate and more interactions are expected at redshifts 
$0.3\lesssim z\lesssim1$ \citep[e.g.][]{KB01}. With the availability of large
telescopes, it is now feasible to conduct spatially resolved spectroscopy of
the faint galaxies at these redshifts to observationally test these
predictions. Therefore, we have performed a large campaign at the VLT targeting
seven distant rich clusters with $0.3\lesssim z\lesssim0.6$. The clusters were
chosen from a very limited list with existing HST/WFPC2 imaging (mainly the core 
regions) at the time the project started (1999) and that are accessible with the 
VLT. \mseins\ has no HST imaging but was included since it was imaged
extensively during FORS science verification time. The main goal is to
derive the two-dimensional internal kinematics of disk galaxies from
emission lines. In combination
with measurements of starformation rates, luminosities and structural 
parameters we aim at disentangling the
effects of different interaction processes and find out about their respective
actual effectiveness and importance for galaxy evolution.

In this Letter, we present the results for the first three clusters of our survey:
\mseins\ ($z=0.30$), \cldrei\ ($z=0.42$), and \clvier\ ($z=0.51$).
They were observed with FORS1 in MOS mode, while the other four clusters 
(Cl\,0016$+$1609, MS\,0451.6--0305, ZwCl\,1447.2$+$2619 \& MS\,2137.3--2353)
had MXU spectroscopic observations with FORS2 requiring different reduction techniques
that will be presented in future papers.
While in \datapaper\ \citep{JZBHM03} we give all the data that can be deduced from 
each single spectrum, we analyze in this Letter only the late--type galaxies which
exhibit spatially resolved emission.
One setup in the MOS mode of FORS1 provides 19 individual slitlets. For each cluster 
two setups have been designed with different rotation angles of the instrument.
Using grism 600R and slitwidths of 1\arcsec, the spectra have a dispersion of 
$\sim$1.08\,\AA/pixel, a spectral resolution of $R\approx1200$ and typical
wavelengths of $\lambda\lambda\approx5200-7400$\AA.
In standard configuration FORS has a
field of view of $6.8\arcmin\times6.8\arcmin$ with a spatial scale of
0.2\arcsec/pixel. To achieve our signal--to--noise requirements of 
$S/N\gtrsim5$ in the emission lines of an $R\lesssim23$ galaxy, the total 
integration time was set to $\sim$2\,hrs. Seeing conditions ranged between 
0.7 and 1.3\arcsec\ FWHM. 

Spatially resolved velocity profiles $V_{\rm rot}(r)$ were determined from
either the [O\,{\sc ii}]3727, H$\beta$ or [O\,{\sc iii}]5007 emission line. 
In eight cases, two lines with sufficient $S/N$ were visible which then were treated
separately yielding consistent results.
The spectral profile of an emission line was fitted by a Gaussian (in case of 
[O\,{\sc ii}] by a double Gaussian) after applying a median filter window of 
typically 0.6\arcsec\ to enhance the $S/N$ stepping along the spatial axis.
Since the apparent disk sizes of spirals at intermediate redshifts are
only slightly larger than the slitwidth (1\arcsec), the slit covers a 
substantial fraction of the two--dimensional velocity field. Therefore, the
spectroscopy is an integration perpendicular to the slit's 
spatial axis. Because of this effect, the maximum rotation velocity 
$V_{\rm max}$ cannot be determined ``straightforward'' from the observed 
rotation. As described in \citet{ZBFJN02} and \citet{BZFBS03}, 
we overcome this problem by simulating such longslit spectroscopy of each galaxy 
individually. 
In short, a two-dimensional velocity field is created assuming a specific 
rotation law that is weighted by the galaxy's luminosity profile and convolved to
match the seeing at the time of our observations. 
Taking into account the galaxy's inclination, position angle, and disk scalelength,
a synthetic rotation curve is generated with $V_{\rm max}$ as the only remaining free
parameter. $V_{\rm max}$ is then determined by matching the synthetic to the
observed velocity profile.
For the galaxies analyzed here, we used for the model rotation curve a simple 
parameterization with a linearily rising inner part that turns over 
into a flat outer part. But as is demonstrated by \citet{BZFBS03}, 
$V_{\rm max}$ is hardly changed when the Universal Rotation Curve by 
\citet{PSS96} is used instead.
The inclinations, position angles and scalelengths were derived via
2-D $\chi^2$--fits of exponential disks to the galaxies' profiles in
FORS images. The FWHM was 0.59\arcsec\ (\mseins), 0.77\arcsec\ (\clvier) and 
1.0\arcsec\ (\cldrei), respectively. 
The fits accounted for the PSF. Spirals with too low an inclination 
($i\lesssim20^\circ$) have not been used for the $V_{\rm max}$ derivation.

Galaxy luminosities were derived from total magnitudes of
FORS images in the $V$ (\mseins) or $I$ (\clvier\ and \cldrei) band,
respectively, as measured with SExtractor \citep{BA96}. Observed magnitudes were 
corrected for Galactic (from \citet{SFD98})
and intrinsic extinction (following \citet{TF85})
and transformed to restframe Johnson $B$ according to their spectrophotometric type
using model SEDs corresponding to Sa, Sb, Sc and Sdm, and calculated for a 
flat $\Omega_\lambda=0.7$ cosmology ($H_0=70$\,km\,s$^{-1}$\,Mpc$^{-1}$).
The average overall error in the photometry is estimated to be $\le0.2^m$.

\section{Kinematics of cluster spirals}

\begin{figure*}
\centerline{\psfig{file=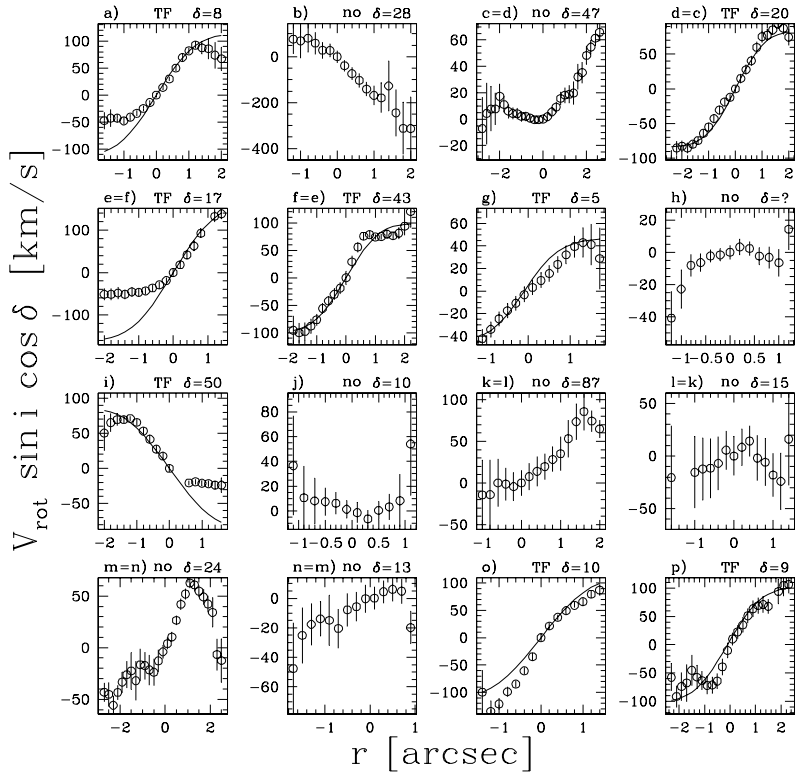,width=11.5cm}}
\caption{\scriptsize \addtolength{\baselineskip}{-3pt}
\label{fig:RC} Spatially resolved velocity profiles for 12 disk 
galaxies in the cluster \mseins\ at $z=0.30$. Note, panels c\&d, e\&f, k\&l 
and m\&n represent the same galaxy, but observed with different slit 
positions (the rotation angles of the two MOS masks differed by 67$^\circ$).
Panels a--p are ordered according to the distance of the galaxies from the 
cluster center. Values of the maximum rotation velocity could be determined 
for cases a, d, e, f, g, i, o \& p indicated by the label ``TF''.
These galaxies enter the TFR shown in Fig.\,2. In all other cases, the kinematics 
are too disturbed (for c \& k this may be due to a large mismatch angle $\delta$).
For the ``TF'' cases, the solid lines denote the \textit{projected} model curves
that were produced from the best fitting 2-D velocity fields simulating the
observational circumstances. Note, that the maximum values of these fits are not
the intrinsic $V_{\rm max}$ values needed for Fig.\,2.
}
\end{figure*}

The MOS mode of FORS1, which was the only multiplex technique available for
the first observations of our campaign, has some disadvantages for the
spectroscopy of spiral galaxies in clusters, which are removed by the MXU mode
of FORS2 that we could use for the other four clusters of our survey.
Firstly, the 19 slitlets of fixed length ($\sim22\arcsec$) have only 
one degree of freedom for placing a slitlet onto an object. 
This leads to somewhat inefficient coverage 
of cluster member candidates and many slitlets must be filled by strongly
relaxing the ideal selection criteria. Secondly, once a certain rotation of 
the field is chosen, all the 19 slitlets have the same orientation on the sky.
Ideally, the slit should be placed along the major axis of a galaxy
to probe its rotation around the center. Although we have targeted each cluster 
with two different setups, the deviation $\delta$ between slit angle and 
position angle was rather large in some cases leading to geometric distortions of 
the observed velocity profile that could not be corrected for. These galaxies are
not used in the further analysis of the internal kinematics here but are valuable
ingredients for future studies of e.g. starformation rates or structural properties
of cluster members.

As is specified in more detail in \datapaper, our method of selecting cluster 
spiral candidates was different for each cluster due to the limited information
of published studies. The most comprehensive source was available for \mseins,
for which we exploited a catalog of $\sim80$ cluster members with published 
spectral types \citep{YEMAC98}. A list of candidates in \clvier\ was prepared 
by comparing optical--nearinfrared colors by \citet{SEDHP02}\footnote{kindly
  provided to us by A. Stanford well in advance of their actual publication} 
to evolutionary stellar population models of an updated version of 
\citet{BC93}. For \cldrei, we mainly utilized a spectroscopic catalog of 
\citet{DG92}. MOS slitlets that could not be filled by a galaxy from our
input lists were placed on objects selected according to their structural 
appearance and magnitudes as measured on our FORS images. If no suitable candidate 
for a late--type galaxy was available, the slitlet was placed on an elliptical 
candidate since understanding the phenomenon of galaxy transformation requires
the analysis of the whole galaxy population of a cluster.

Redshifts and spectral types could be determined for
12/13 spiral/elliptical members of \mseins\ (plus 4/2 S/E field galaxies),
8/1 spiral/elliptical members of \clvier\ (plus 11/6 S/E field galaxies), and
10/7 spiral/elliptical members of \cldrei\ (plus 15/0 S/E field galaxies).
While the early--type galaxies will be discussed in a future paper, we here
concentrate on the emission--line galaxies. As was pointed out by \citet{Verhe01}
only galaxies with a rotation curve that rises in the inner region and then
clearly turns into a flat part should be used for a Tully--Fisher diagram.
In such a case, the measured $V_{\rm max}$ is representative for the
influence of the dark matter halo on the galaxy's kinematics and is indicative
for the total dynamical mass of the galaxy. For our three clusters, we were able
to determine $V_{\rm max}$ for 7/5/1 different member galaxies, respectively, (and
1/5/1 field spirals). 
The remaining cluster spirals either have too low $S/N$ of their emission 
lines to spatially analyse the internal kinematics (in 0/0/4 cases),
have too low inclination $i$ or too large mismatch angle $\delta$ (in 1/3/5 cases), 
or exhibit intrinsic distortions (in 4/0/0 cases).

In Fig.\,\ref{fig:RC} we show as an example position--velocity diagrams for
all member galaxies in \mseins\ that have sufficient $S/N$. Of the 12 members, of 
which four were observed twice with different slit angles with respect to 
their major axis, seven exhibit the ``classical'' rotation curve shape rising
in the inner part and turning over to a flat regime (labeled ``TF''). Four members
clearly show disturbed kinematics (panels b, j, l, \& m). The distortion seen in 
panel m (\& n) most probably arises from a bar which is readily visible in the
direct image. Two ``double--hits'' were observed with very big mismatch angles
(panels c \& k) so that their velocity profiles look peculiar.
In one case (panel h) the object fell onto the slitlet by chance and is too weak
(small) to derive its structural parameters.

In the figure, the galaxies are ordered according to their projected distance to
the cluster center (ranging from 285\,kpc to 1230\,kpc). 
There is no trend visible of the RC form as a function of 
clustercentric distance, i. e. distorted velocities are not uniquely tied to
the central region. But since our observations cover only the region within 
the virial radius of the cluster ($\sim1880$\,kpc), 
this is in accordance with dynamical 
models in which the galaxy population of a cluster is well-mixed within that 
region \citep{BNM00}.
In particular, we most probably do not have any new arrivals from the field
in our sample.

\section{The Tully--Fisher relation}

In Fig.\,2, we present the Tully--Fisher diagram for the distant
cluster galaxies. Only those galaxies for which $V_{\rm max}$ could be
determined enter the plot. The ordinate gives the restframe absolute Johnson
$B$-band magnitudes. We also show the position of those field galaxies that 
were serendipitously observed in the cluster fields and that have $0.4<z<0.6$
(3 spirals at $z=0.61$ may be members of a background cluster behind \clvier).
For comparison, we include the distant field galaxies from the F{\sc ors} Deep 
Field \citep{HAGJS03} which were observed with exactly the same instrumental setup 
\citep{ZBFJN02,BZFBS03}. The linear bisector fit to this sample is flatter than 
the slope fitted to the local sample of \citet[][PT92]{PT92}, which is given
with its $\pm3\sigma$ deviations.

The distant cluster spirals are distributed very much alike the field population 
that covers similar cosmic epochs ($0.1<z<1$ with the bulk of galaxies at
$z\approx0.3$ and $\approx0.6$).
No significant deviation from the distant field TFR is visible and the cluster
sample has not any increased scatter, but the low number of cluster members 
prohibits any quantitative statistical analysis. Nevertheless, we can conclude
that the mass--to--light ratios of the observed distant cluster spirals cover the
same range as the distant field population indicating that their stellar 
populations were not dramatically changed by possible clusterspecific interaction
phenomena. In particular, we do not detect any significant overluminosities as
would be expected in the $B$ band if strong starbursts had occured in the recent 
past of the examined cluster galaxies.
With respect to the Tully--Fisher relation obeyed by local galaxies
(e.g. PT92), our cluster sample follows the same trend as the
FDF galaxies. Since mostly only the bright galaxies made it into the TF diagram, 
the cluster members occupy a region where no significant luminosity evolution
is visible.

\centerline{\psfig{file=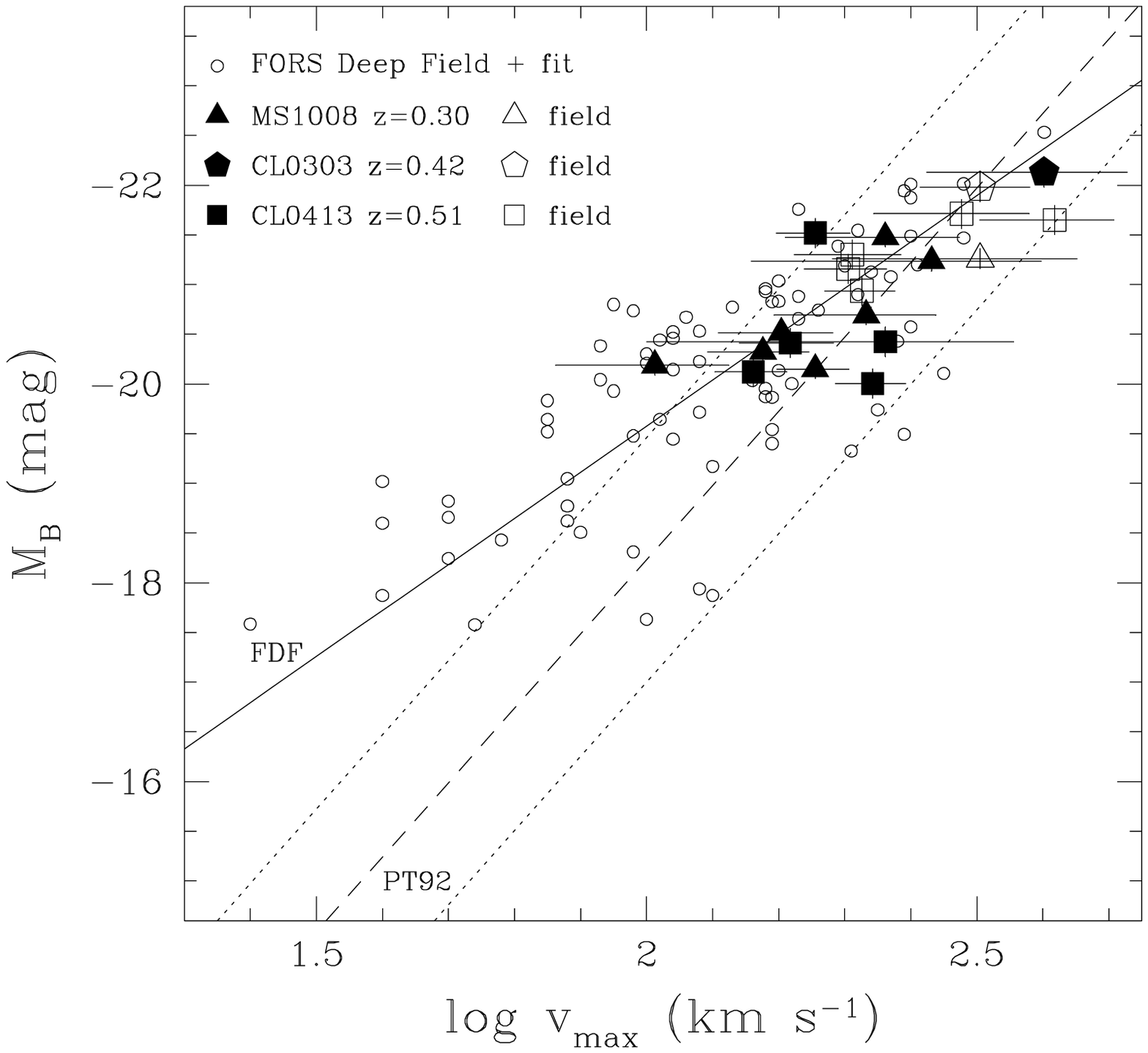,width=3.0in}}\nopagebreak
\smallskip
\noindent{\scriptsize \addtolength{\baselineskip}{-3pt} {\sc Fig. 2.}---
Tully--Fisher diagram of cluster spirals 
in \mseins\ (filled triangles), in \clvier\ (filled squares), and \cldrei\
(filled diamond).
Also shown are seven field objects (open symbols) that were serendipitously 
observed. In comparison to the F{\sc ors} Deep Field sample of 77 field 
galaxies \citep{ZBFJN02,BZFBS03} with $0.1<z<1$ (small open circles), the cluster 
galaxies are similarly distributed and do not deviate significantly from
the linear fit to the FDF sample (solid line).
The cluster members follow the same trend with respect to the local TFR
(the fit $\pm3\sigma$ to the \citet{PT92} field sample is given)
as the distant field galaxies with the brightest galaxies exhibiting the
smallest luminosity evolution.}
\bigskip

But we emphasize that this conclusion is true only for those objects that
enter the TF diagram. Since a significant fraction of our cluster galaxies can not be
used for a TF analysis due to their distorted kinematics the above conclusions
are not generally valid for the whole cluster sample.
The objects with intrinsically peculiar velocity curves may actually be subject to 
ongoing or may have experienced recent interactions. Such processes most probably
also influence the stellar populations of a galaxy changing its integrated 
luminosity as well. Tidal interactions for example could distort the spiral arms while
inducing starbursts at the same time leading to enhanced luminosities.
Since we do not know where the galaxies with peculiar kinematics lie in the TF plane,
it is not possible to decide whether a particular galaxy has an increased or decreased
luminosity. Overall, these galaxies span a very similar range in apparent magnitudes 
to the TF cluster members.


Exploring the spatially resolved kinematics of disk galaxies in distant clusters 
has become feasible only recently. \citet[][MJ03]{MAHJH03} studied a sample of 7 
spiral galaxies in the cluster MS\,1054.4--0321 at $z=0.83$ with FORS at the VLT
in a similar configuration as our own spectroscopy. 
Compared to a number of field spirals at corresponding redshifts, which were 
observed at the same time, they find that the cluster members have brighter $B$ 
luminosities by $\sim0.5-1^m$ ($\sim1.5-2\sigma$ significance) for their
rotational velocities. The difference to the average brightening of our
cluster members is hardly significant and may be due to a combination of
low-number statistics and systematic deviations. But we also can not rule out that
the differences are real and may be connected to the higher redshift of MS\,1054
or other characteristics of that cluster.

\citet{Metev03} et al. examined galaxies in the cluster Cl\,0024$+$1654 
($z=0.40$) with the Keck 10m-telescope. The ten galaxies that appear in their
TF diagram have a larger scatter than the local PT92 sample, but show no evidence
for an evolution of the zero-point. The authors argue that processes acting on
the cluster galaxies some time before the lookback time of the observations 
involving either starbursts or a truncation of star formation may have caused a
decreased or increased mass--to--light ratio, respectively, which is manifested in the
increased scatter.

In a future paper we will present our analysis for the other four clusters of
our campaign. With more cluster member galaxies both in the TF diagram and those
with peculiar kinematics, we may hopefully be able to quantitatively investigate
the galaxy evolution in rich clusters and to give also statistical tests.

\acknowledgments
We acknowledge the thorough comments by the referee.
We are very grateful to Drs. S. Wagner (Heidelberg), U. Hopp (M\"unchen) and
R. H. Mendez (Hawaii) for performing part of the observations and thank ESO and 
the Paranal staff for efficient support. 
We also thank the PI of the FORS project, Prof. I. Appenzeller (Heidelberg), and
Prof. K. J. Fricke (G\"ottingen) for providing guaranteed time for our project.
We also acknowledge fruitful discussions with Drs. B. Milvang--Jensen (MPE Garching)
and M. Verheijen (Potsdam).
This work has been supported by the Volkswagen Foundation (I/76\,520)
and the Deutsche Forschungsgemeinschaft (Fr 325/46--1 and SFB 439).



\clearpage


\clearpage






\end{document}